\documentclass{elsarticle}

\usepackage{amsmath, amssymb}
\usepackage{mathtools}
\usepackage{braket}
\usepackage{siunitx}
\usepackage{textcomp}
\usepackage{braket}

\usepackage{math}
\usepackage{color}

\usepackage{hyperref}

\begin{document}

\begin{frontmatter}
\title{Theoretical analysis of the electron bridge process in $^{229}$Th$^{3+}$}
\author[PTB,TUB]{R. A. M\"uller\corref{corauth}}
\cortext[corauth]{Corresponding author}
\ead{robert.mueller@ptb.de}
\author[HIJ]{A. V. Volotka}
\author[HIJ,FSU]{S. Fritzsche}
\author[PTB,TUB]{A. Surzhykov}
\address[PTB]{Physikalisch-Technische Bundesanstalt, D-38116 Braunschweig, Germany}
\address[TUB]{Technische Universit\"at Braunschweig, D-38106 Braunschweig, Germany}
\address[HIJ]{Helmholtz-Institute Jena, D-07743 Jena, Germany}
\address[FSU]{Theoretisch-Physikalisches Institut, Friedrich-Schiller-Universit\"at Jena, D-07743 Jena, Germany}
\begin{abstract}
We investigate the deexcitation of the $^{229}$Th nucleus via the excitation of an electron. Detailed calculations are performed for the enhancement of the nuclear decay width due to this so called \emph{electron bridge} (EB) compared to the direct photoemission from the nucleus. The results are obtianed for triply ionized thorium by using a B-spline pseudo basis approach to solve the Dirac equation for a local $x_\alpha$ potential. This approach allows for an approximation of the full electron propagator including the positive and negative continuum. We show that the contribution of continua slightly increases the enhancement compared to a propagator calculated by a direct summation over bound states. Moreover we put special emphasis on the interference between the direct and exchange Feynman diagrams that can have a strong influence on the enhancement.
\end{abstract}
\end{frontmatter}
\section{Introduction}
\label{sec:introduction}
Because of its extremely low lying first excited metastable state the $^{229}$Th nucleus has been object of intense investigation during the last decades \cite{kroger_features_1976, reich_energy_1990, guimaraes-filho_energy_2005, beck_energy_2007, karpeshin_subthreshold_1996, karpeshin_3.5-ev_1999, karpeshin_excitation_2015, porsev_effect_2010, porsev_electronic_2010, porsev_excitation_2010}. Due to its extremely narrow linewidth the transition between this isomeric and the nuclear ground state is planned to be the working transition of the future nuclear clock \cite{peik_nuclear_2003, peik_nuclear_2015, von_der_wense_direct_2016}. This clock will be very important for benchmarking existing atomic clocks and might help to provide a new optical frequency standard \cite{peik_prospects_2008}. Moreover this clock will be accurate enough to test predictions for the fifth force and the time variation of fundamental constants \cite{berengut_proposed_2009, rellergert_constraining_2010}. Triply charged thorium is a good candidate to study in particular because it has been successfully laser cooled and was already studied in ion trap experiments \cite{campbell_multiply_2009}.

So far the exact energy of the low lying nuclear resonance in $^{229}$Th remains unknown. Therefore many proposals have been put forward how to address this level \cite{palffy_theory_2006, porsev_excitation_2010, karpeshin_electron_2002}. Meanwhile the majority of these scenarios have been realized in experiments aiming for a precise determination of the resonance energy \cite{kroger_features_1976, reich_energy_1990, guimaraes-filho_energy_2005, von_der_wense_direct_2016, okhapkin_observation_2015, yamaguchi_experimental_2015, peik_nuclear_2003}. A key property here is the lifetime of the isomeric state that is strongly influenced by the electronic environment, inter alia because of the electron bridge (EB) process \cite{porsev_effect_2010}. In this process the $^{229}$Th nucleus does not decay via emission of a photon but by exciting the electron shell. We present here an approach to analyze the influence of the EB onto the decay width of the first excited state of the $^{229}$Th nucleus. Our approach extends the analysis shown in Ref. \cite{porsev_effect_2010} by two means. First our method allows to include the experimental energies of the important transitions in Th$^{3+}$ and does not necessarily rely on many-electron calculations that are not always in good agreement with the measured spectra. Second we approximate the full electron propagator, including the coupling to the positive and negative continuum, in contrast to Ref. \cite{porsev_effect_2010}, where a direct summation over bound states is performed.

After we have sketched our theory in Sec. \ref{sec:theory} we will present results in Sec. \ref{sec:results}, where we will discuss the role of the positive an negative continuum in the electron propagator as well as interference effects between the participating quantum processes. Natural units ($\hslash=c=1$) are used throughout this work.
\section{Theory}
\label{sec:theory}
%
%
\begin{figure}
\centering
\includegraphics[width=\textwidth]{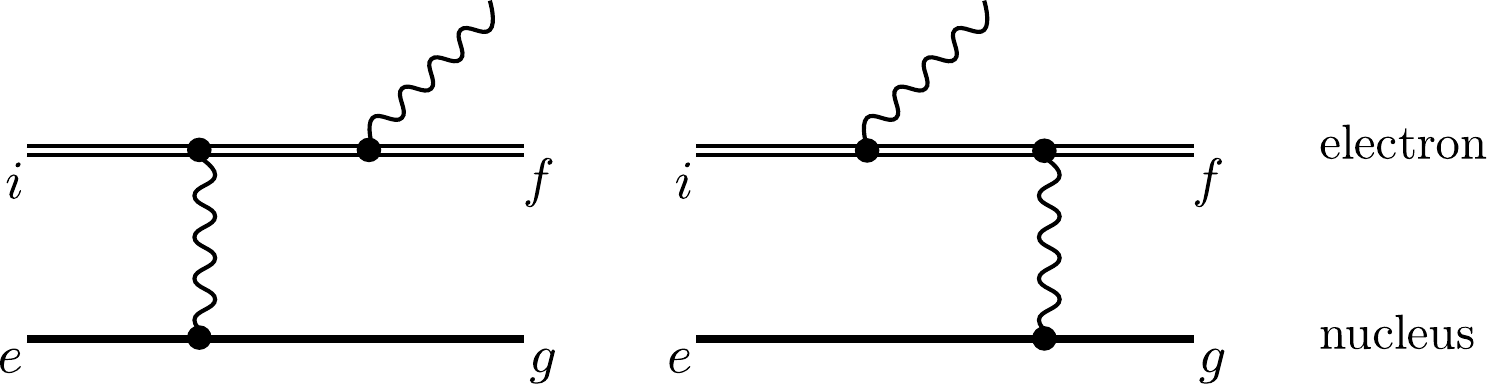}
\caption{Feynman diagrams for the electronic bridge process. $e$ and $g$ are the first excited and the ground state of the nucleus, $i$ and $f$ the initial and final electronic states.\label{fig:diagrams}}
\end{figure}
\begin{figure}
\centering
\includegraphics[width=\textwidth]{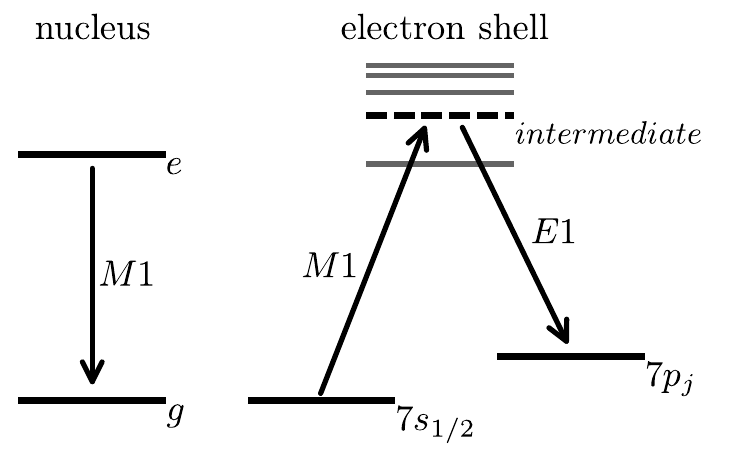}
\caption{Sketch of the electronic bridge process for the deexcitation of the nucleus via the $M1$-excitation of a $7s_{1/2}$ electron to a (virtual) intermediate state. In a second step the intermediate state decays via an $E1$ electric dipole transition to a $7p_j$ state. \label{fig:process}}
\end{figure}
The EB is described by two Feynman diagrams shown in Fig. \ref{fig:diagrams}. The \emph{direct diagram} describes the process where the nuclear isomeric state $|e\rangle$ decays to the ground state $|g\rangle$ via exchanging a virtual photon with the electron shell from which, in turn, a real photon is emitted. In the \emph{exchange diagram} the real photon is emitted before the deexcitation of the nucleus takes place. It is known experimentally that the $|e\rangle\rightarrow|g\rangle$ transition in $^{229}$Th  is of magnetic dipole type ($M1$). As shown in Fig. \ref{fig:process} this transition leads to an $M1$ excitation of the electron to a, possibly virtual, intermediate state. We will employ the dipole approximation for the emitted real photon so that the transition from the intermediate to the final electronic state is of electric dipole type ($E1$).

In order to quantify the influence of the EB on the total decay width of the $^{229}$Th nucleus we will derive an expression for the so called \emph{enhancement factor} $\beta$. It is defined by the ratio between the width $\Gamma_{EB}$ of the EB process, compared to the width of the direct $M1$ photo decay of the nucleus:
\begin{equation}
\beta = \frac{\Gamma_{EB}}{\Gamma_N(M1, e\rightarrow g)}
\label{eq:beta_def}
\end{equation}
The spontaneous decay width $\Gamma_N(M1, e\rightarrow g)$ has not been measured yet. Only a theoretical estimate has been given \cite{dykhne_matrix_1998}. We will see later, that $\beta$ is especially useful because it is largely independent on $\Gamma_N(M1, e\rightarrow g)$. This makes the enhancement factor $\beta$ a convenient quantity for the comparison of different theories and scenarios.

The calculation of the EB-width $\Gamma_{EB}$ can be traced back to an evaluation of the transition amplitudes. We apply the Feynman rules to the diagrams shown in Fig. \ref{fig:diagrams} to obtain expressions for these. Assuming that the angular momentum projections of the initial and final nuclear and electronic states are not observed, we obtain for $\Gamma_{EB}$:
\begin{equation}
\begin{aligned}
\Gamma_{EB} =& \left(\frac{\omega}{\omega_N}\right)^3\frac{\alpha\Gamma_N(M1, e\rightarrow g)}{4\pi(2j_i+1)}\\
\times&\left[\sumint_n \frac{1}{2j_n +1}\left(\left|\frac{\braket{f||\hat{D}||n}\braket{n||\hat{T}_1||i}}{\omega_{in} + \omega_N}\right|^2 + \left|\frac{\braket{f||\hat{T}_1||n}\braket{n||\hat{D}||i}}{\omega_{fn} - \omega_N}\right|^2\right)\right.\\
&+\left.\sumint_{nn'}(-1)^{j_n+j_{n'}} \left\lbrace
\begin{matrix}
j_i&L&j_n\\
j_f&1&j_{n'}
\end{matrix}
\right\rbrace\frac{\braket{f||\hat{D}||n}\braket{n||\hat{T}_1||i}\braket{f||\hat{T}_1||n'}^*\braket{n'||\hat{D}||i}^*}{(\omega_{in}+\omega_N)(\omega_{fn'}-\omega_N)}\right],
\end{aligned}
\label{eq:eb_width}
\end{equation}
where the electronic states $|i\rangle$ have angular momentum $j_i$ and energy $\epsilon_i$. The energy of the emitted photon is $\omega=\epsilon_i-\epsilon_f+\omega_N$, $\alpha$ is the fine structure constant and $\omega_N$ the energy splitting between the nuclear ground and first excited state. The electron propagator is represented by summing and integrating over all possible intermediate states $n$. The energy splitting between these intermediate and the initial or final electronic states is labeled by $\omega_{i/fn}$. We can see that $\Gamma_{EB}$ increases drastically if the resonance condition $\omega_{in}+\omega_N=0$ is fulfilled. In this resonant case it is important to include the width of the atomic states to resolve the divergence of the denomiators in Eq. \eqref{eq:eb_width}. Moreover $\Gamma_{EB}$ scales linearly with $\Gamma_N(M1,e\rightarrow f)$, which cancels with the denomiator of Eq. \eqref{eq:beta_def} and makes $\beta$ independent of the width of isomeric nuclear state.

The transition operators in Eq. \eqref{eq:eb_width} are the dipole operator $\hat{D}$ describing the photon emission from the electron shell and the operator $\hat{T}_{1M}$ which is part of the virtual photon exchange in the electron nucleus interaction (cf. Fig. \ref{fig:diagrams}). Following similar steps to Refs. \cite{plunien_nuclear_1991, tkalya_excitation_2014, volotka} find for $\hat{T}_{1M}$:
\begin{equation}
\hat{T}_{1M}=-e\sqrt{\frac{4\pi}{3}}\frac{[\hat{\vec{r}}\times\vec{\alpha}]_M}{r^3},
\end{equation}
where $\vec{\alpha}$ is the vector of Dirac matrices and $e$ is the electron charge.
\section{Computational details}
%
%
After we have derived an expression for the EB-width $\Gamma_{EB}$ \eqref{eq:eb_width} we need to obtain a basis set for all (bound and continuum) electronic states to evaluate the matrix elements and the sum and integral in Eq. \eqref{eq:eb_width}. In order to to this, we restrict ourselves here to the single active electron (SAE) approximation, which is well justified since Th$^{3+}$ has only one valence electron. In order to generate the wave function for this valence electron, we solve the Dirac-Equation for a $x_\alpha$ potential, that is assembled from the Coulomb potential $V_{nuc}(r)$ of the extended nucleus and the static potential generated by the other electrons:
\begin{equation}
V_{x_\alpha}(r)=V_{nuc}(r) + \alpha\int_0^\infty dr' \frac{\varrho (r')}{\mathrm{max}(r,r')} - x_\alpha\frac{\alpha}{r} \left(\frac{81}{32\pi^2} r \varrho (r)\right)^{\frac{1}{3}},
\label{eq:ks potential}
\end{equation}
where $\varrho(r)$ is the electron density obtained by means of the Dirac-Hartree-Fock method. For the usual Kohn-Sham potential the prefactor of the last term is $x_\alpha=\frac{2}{3}$. In our case we take $x_\alpha$ as a free parameter and vary it so that the corresponding binding energies accuratly match the transition energy  between the initial and final electronic state. Following Refs. \cite{sapirstein_use_1996, shabaev_dual_2004} we construct for this potential a finite pseudo basis set consisting of B-spline wave functions that are solutions of the Dirac equation. This reduces the infinite sum and integral in Eq. \eqref{eq:eb_width} to a finite sum and allows for a very good approximation of the electron propagator \cite{volotka_interelectronic-interaction_2011, volotka_many-electron_2016}.  Now by plugging Eq. \eqref{eq:eb_width} into Eq. \eqref{eq:beta_def}, we can calculate the enhancement factor $\beta$.

\section{Results and discussion}
\label{sec:results}
%
%
\begin{table}
\centering
\caption{Comparison between the experimental spectrum \cite{klinkenberg_spectral_1988} and the spectra obtained using our screening potentials \eqref{eq:ks potential} varied to match the transition energy between $7s_{1/2}\rightarrow 7p_{1/2}$ and $7s_{1/2}\rightarrow 7p_{3/2}$, respectively. All energies are given in $\mathrm{eV}$ with respect to the ionization threshold. \label{tab:comp_energ}}
\begin{tabular}{c|ccc}
State & experiment \cite{klinkenberg_spectral_1988} & fit $7s_{1/2} \rightarrow 7p_{1/2}$ & fit $7s_{1/2} \rightarrow 7p_{3/2}$\\
\hline\hline
$5f_{5/2}$ & $-28.811$ & $-36.403$ & $-35.142$\\
$5f_{7/2}$ & $-28.275$ & $-35.508$ & $-34.271$\\
$6d_{3/2}$ & $-27.672$ & $-28.843$ & $-28.238$\\
$6d_{5/2}$ & $-27.015$ & $-27.859$ & $-27.287$\\
$7s_{1/2}$ & $-25.944$ & $-26.094$ & $-25.808$\\
$7p_{1/2}$ & $-21.343$ & $-21.493$ & $-21.301$\\
$7p_{3/2}$ & $-19.754$ & $-19.766$ & $-19.621$\\
$8s_{1/2}$ & $-13.980$ & $-14.133$ & $-14.038$\\
$7d_{3/2}$ & $-13.972$ & $-14.170$ & $-14.074$\\
$7d_{5/2}$ & $-13.756$ & $-13.943$ & $-13.846$\\
$6f_{5/2}$ & $-13.033$ & $-13.163$ & $-13.067$\\
$6f_{7/2}$ & $-12.964$ & $-13.099$ & $-13.005$\\
$8p_{1/2}$ & $-12.133$ & $-12.225$ & $-12.147$\\
$8p_{3/2}$ & $-11.469$ & $-11.480$ & $-11.415$\\
$9s_{1/2}$ & $-8.8837$ & $-8.9192$ & $-8.8756$
\end{tabular}
\end{table}
Before we present calculations for the enhancement factor $\beta$ \eqref{eq:beta_def}, we want to convince ourselves that the approximations we made to obtain the electron wavefunctions are valid. Therefore we will use our potential to calculate the spectrum of Th$^{3+}$ and compare these calculations with experimental data. Th$^{3+}$ has one electron above a closed radon core,  the ground state configuration is $[$Rn$]5f^1$. Thus, for brevity, we can name the ionic configurations by the state of the valence electron. Many experiments are performed not using the ground but excited states of Th$^{3+}$. The $7s_{1/2}$ state is of particular interest here because it has resonances near the expected nuclear excitation energy. Due to the dipole transition matrix elements in Eq. \eqref{eq:eb_width}, the final state has to be of opposite parity and we will restrict ourselves here to a final $7p_j$ state. Therefore we vary the parameter $x_\alpha$ in Eq. \eqref{eq:ks potential} to match the calculations with the experimental values \cite{klinkenberg_spectral_1988} for the transitions $7s_{1/2}\rightarrow 7p_{1/2}$ and $7s_{1/2}\rightarrow 7p_{3/2}$. After variation we obtain $x_\alpha=1.06$ for the $7p_{1/2}$ and $x_\alpha=1.01$ for the $7p_{3/2}$ case. With these potentials we can now compare the calculated spectra with the measured level energies as shown in Tab. \ref{tab:comp_energ}. It can be seen from the table that our agreement with the experimental energies is always better than $2\%$ for states above the $6d$-shell. Calculations for the Th$^{3+}$ spectrum that we performed using the GRASP2k package \cite{jonsson_new_2013} help us to explain the strong disagreement for the lower lying levels. These calculations show us that for these levels correlation effects play an important role which are not covered by the approximation \eqref{eq:ks potential}.

\begin{figure}
\centering
\includegraphics[width=\textwidth]{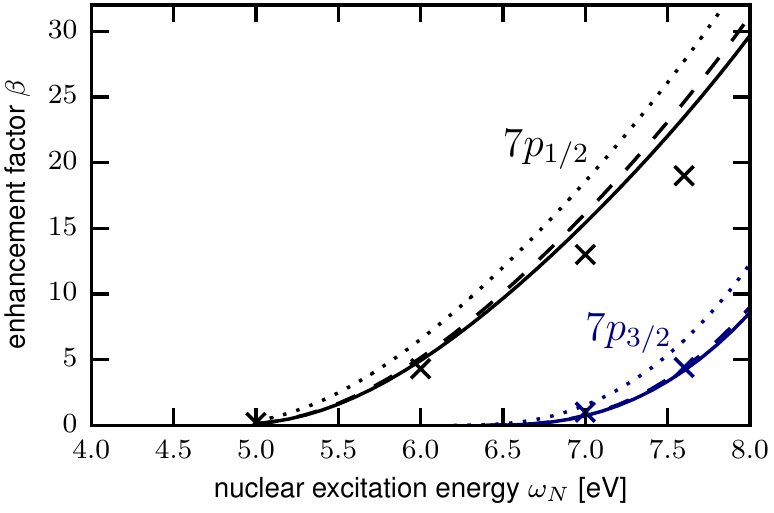}
\caption{Calculations for the enhancement factor $\beta$ as a function of $\omega_N$ for two different transition paths (i) $7s_{1/2}\rightarrow 7p_{1/2}$ (black lines and symbols, upper set) and (ii) $7s_{1/2}\rightarrow 7p_{3/2}$ (blue lines and symbols, lower set). The crosses are values published in Ref. \cite{porsev_effect_2010} and the solid lines correspond to results obtained with our screening potential fitted to the energies and the basis set shown in Ref. \cite{porsev_effect_2010}. The dashed lines correspond to results obtained with the same potential but the full basis set, while the dotted lines are results where we varied the potential to match the experimental transition energies. \label{fig:comp_pors}}
\end{figure}
Above we have shown that our approximation provides good results for the desired transition from the initial $7s_{1/2}$ to the final $7p_j$ state. This allows us to perform calculations for the enhancement factor $\beta$. Because the nuclear excitation energy $\omega_N$ is unknown we take it as a free parameter and evaluate $\beta$ as a function of $\omega_N$ as shown in Fig. \ref{fig:comp_pors}. Generally it can be seen in the figure that $\beta$ increases towards higher energies. This is due to the closeby $7s_{1/2}\rightarrow 8s_{1/2}$ resonance. Therefore the enhancement factor $\beta$ is larger than one over a wide energy range, which means that the deexcitation via the EB process is more probable than the emission of a photon from the nucleus.

In order to draw a comparison to previous results, we modified our approach to mimic the theory put forward by Porsev and Flambaum \cite{porsev_effect_2010}. Therefore we (i) varied our potential to match the energies published in Ref. \cite{porsev_effect_2010} ($x_\alpha=1.12$ for $7s_{1/2}\rightarrow 7p_{1/2}$, $x_\alpha=1.07$ for $7s_{1/2}\rightarrow 7p_{3/2}$) and (ii) truncated our basis to the same set of states used by Porsev and Flambaum. The results of these calculations are shown as solid lines in Fig. \ref{fig:comp_pors} together with the set of values presented by Porsev and Flambaum \cite{porsev_effect_2010} (crosses) for both final states $7p_{1/2}$ (upper black set) and $7p_{3/2}$ (lower blue set). It is seen that we achieve a very good agreement with these previous calculations, especially for the $7p_{3/2}$ final state. But it is important to note, that due to the incomplete set of intermediate states the theory is not gauge invariant. The dashed lines in contrast show the results for the same potential but the full B-spline pseudo basis. In order to check the gauge invariance of this approximation we performed calculations in length and velocity gauge. These results turn out to agree up to the order $10^{-5}$. Compared to the calculations with the truncated basis set $\beta$ is slightly increased if the full set of B-splines is used. The dotted lines show results again for the complete pseudo basis and a potential where the binding energies are matched to the experimental values. The fact that $\beta$ is again larger in this case is mainly due to the fact that the experimental energy of the nearest resonance, where $\omega_N=\epsilon_{8s_{1/2}}-\epsilon_{7s_{1/2}}$, is about $0.3\mathrm{eV}$ lower than calculated in Ref. \cite{porsev_effect_2010}. This shows us that even in the regime where $\omega_N$ is far from an electronic resonance it is important to have an accurate representation of the electronic spectrum.

\begin{figure}
\centering
\includegraphics[width=\textwidth]{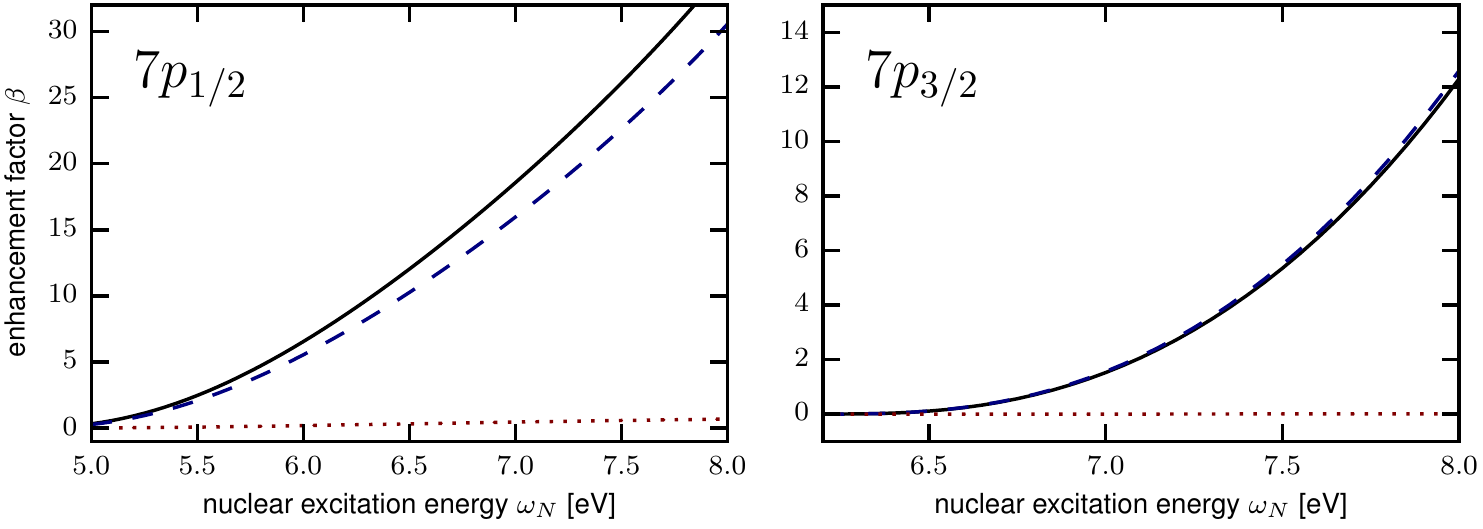}
\caption{Calculations for the enhancement factor $\beta$ for the initial state $7s_{1/2}$ and the final state $7p_{1/2}$ (left panel) and $7p_{3/2}$ (right panel). The solid lines are the full result with the potential fitted to the experimental spectrum just as shown already in Fig. \ref{fig:comp_pors}. The blue dashed lines show $\beta$ calculated only for the direct and the red dotted lines only for the exchange diagram. \label{fig:betascan}}
\end{figure}
All results shown above in Fig. \ref{fig:comp_pors} were obtained including the contributions of both the direct and the exchange diagram (cf. Fig. \ref{fig:diagrams}). In Fig. \ref{fig:betascan} we show a more detailed investigation of the different contributions to $\beta$ from both diagrams involved. The calculations are performed again for two final states and with the use of the potential fitted to the experimental transition energies as presented in Tab. \ref{tab:comp_energ}. In each panel of Fig. \ref{fig:betascan} the solid black lines correspond to the full results \eqref{eq:eb_width} as already shown in Fig. \ref{fig:comp_pors}. The dashed blue and dotted red lines show $\beta$ calculated only for the direct or the exchange diagram, respectively. It can be seen that in the case of the final state being $7p_{1/2}$ the full result does not correspond to the sum of direct and exchange amplitudes. While the contribution from the exchange diagram is almost zero, the full result is about $10\%$ larger than the calculation from the direct diagram. This increase is due to the interference between the two processes. For the final state $7p_{3/2}$ this inteference effeect is negative but negligible. Conclusively Fig. \ref{fig:betascan} shows us that, however the contribution from the exchange diagram is small, it cannot always be neglected because of strong interference effects.

\section{Summary}
\label{sec:summary}
We have investigated the enhancement of the decay width of the low lying nuclear isomeric state due to the EB process in $^{229}$Th$^{3+}$. Our method allows us to obtain results for this enhancement as a function of the nuclear excitation energy $\omega_N$, where we are able to accuratly match the important transition energies to the experimental values. We have shown that a good representation of the spectrum is important even in the off-resonance regime. Moreover we found out that the enhancement due to the EB is slightly underestimated if the electron propagator is approximated by a direct sum over bound electron states. However the contribution from the exchange diagram has been rightly neglected in some works \cite{porsev_electronic_2010, porsev_excitation_2010} we have shown that in some cases interference effects between the direct and the exchange diagram can have a large influence on the width of the EB.

\section*{Acknowledgements}
The authors are grateful for many useful discussions with Maksim Okhapkin.
RAM acknowledges support of the RS-APS and the HGS-HIRe.

\section*{References}
\bibliography{quellen.bib}
\end{document}